\documentclass[conference,a4paper]{IEEEtran}
\usepackage{amsmath,amssymb,epsfig,color,flushend,epstopdf,algorithm}
\usepackage[noend]{algpseudocode}

\begin{document}
\title{Joint Transmit Power and Relay Two-Way Beamforming Optimization for Energy-Harvesting Full-Duplex Communications}
\author{\IEEEauthorblockN{Alexander A. Okandeji$\stackrel{\S}{,}$ Muhammad R. A. Khandaker$\stackrel{\S}{,}$ Kai-Kit Wong$\stackrel{\S}{,}$ and Zhongbin Zheng$^\ddag$}
\IEEEauthorblockA{$^\S$Department of Electronic and Electrical Engineering\\
University College London\\
Torrington Place, London, WC1E 7JE, United Kingdom\\
E-mail: $\rm alexander.okandeji.13@ucl.ac.uk, m.khandaker@ucl.ac.uk, kai\text{-}kit.wong@ucl.ac.uk$\\
$^\ddag$East China Institute of Telecommunications\\
China Academy of Information and Communications Technology\\
E-mail: $\rm ben@ecit.org.cn$}}

\maketitle
\begin{abstract}
This paper studies the joint optimization problem of two-way relay beamforming, the receiver power splitting (PS) ratio as well as the transmit power at the sources to maximize the achievable sum-rate of a simultaneous wireless information and power transfer (SWIPT) system with a full-duplex (FD) multiple-input multiple-output (MIMO) amplify and forward (AF) relay, assuming perfect channel state information (CSI). In particular, our contribution is an iterative algorithm based on the difference of convex programming (DC) and one dimensional searching to achieve the joint solution. Simulation results are provided to demonstrate the effectiveness of the proposed algorithm.
\end{abstract}

\section{Introduction}
Traditionally, wireless communication systems use a time-division or frequency-division approach to bidirectional communication. This involves dividing the spectral resources into orthogonal component resulting in half-duplex (HD) communication. Recent advances, nevertheless, suggest that full-duplex (FD) communication that allows simultaneous transmission and reception of signal over the same radio channel be possible \cite{exp}. Since radio signals that carry information can also be used as a vehicle for transporting energy, the emergence of FD technology brings a new opportunity for simultaneous wireless information and power transfer (SWIPT) \cite{Alex,Alex_2}. A practical application of SWIPT technology is seen in battery-limited devices such as sensor nodes mounted at some inaccessible or difficult-to-access locations \cite{mimo}, \cite{ruhuul2}.

Recently, much interest has turned to FD relaying in which information is sent from a source node to a destination node through an intermediate FD relaying node. In the literature, relay aided SWIPT systems have been largely considered for HD relaying  \cite{Na}. Recently, the authors in \cite{transG} considered SWIPT in FD relaying where only the relay node works in FD mode. Most recently, \cite{Alex_2} considered SWIPT in FD multiple-input multiple-output (MIMO) relay system and used a power splitting (PS) relaying approach with fixed transmit power at the source nodes. However, the transmit power in FD systems affects the self-interference (SI) and careful optimization is necessary in order to maximize the achievable rate.

In contrast to the existing results, this paper investigates the joint optimization of the transmit power at the source nodes and the two-way relay beamforming matrix for SWIPT with a FD MIMO amplify and forward (AF) relay employing PS, where the achievable sum-rate is maximized subject to energy harvesting and individual power constraints.

{\em Notations}---We use ${\bf X}\in\mathbb{C}^{M \times N}$ to represent a complex matrix with dimension of $M \times N$. Also, we use $(\cdot)^{\dagger}$ to denote the conjugate transpose, while $\mathrm{trace}(\cdot)$ is the trace operation, and $\|\cdot\|$ denotes the Frobenius norm. In addition, $| \cdot|$ returns the absolute value of a scalar, and ${\bf X}\succeq {\bf 0}$ denotes that the Hermitian matrix ${\bf X}$ is positive semidefinite. The expectation operator is denoted by $\mathbb{E}\{\cdot\}.$ We define $\Pi_\mathbf{X} = \mathbf{X}(\mathbf{X}^{\dagger}\mathbf{X})^{-1}\mathbf{X}^{\dagger}$ as the orthogonal projection onto the column space of $\mathbf{X}$; and $\Pi^{\perp}_{\mathbf{X}} = \mathbf{I} - \Pi_\mathbf{X}$ as the orthogonal projection onto the orthogonal complement of the column space of $\mathbf{X}.$

\section{System Model and Problem Formulation}
\subsection{System Model}
\begin{figure}
\begin{center}
\includegraphics[width=8cm]{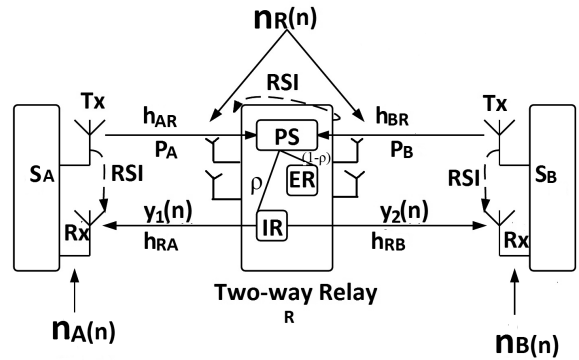}
\caption{The model of the two-way FD SWIPT system.}\label{fig-sys}
\end{center}
\end{figure}

Consider SWIPT in a three-node relaying network consisting of two sources $\rm S_A$ and $\rm S_B$ wanting to exchange information with the aid of an intermediate MIMO AF relay $\mathrm{R},$ as shown in Fig.~\ref{fig-sys}. In this model, the two source nodes $\rm S_{A}$ and $\rm S_{B}$ and the relay $\rm R$ are assumed to operate in FD mode. We also assume that there is no direct link between $\rm S_{A}$ and $\rm S_{B},$ so communication between them must be done through $\mathrm{R}.$ Both $\rm S_{A}$ and $\rm S_{B}$ transmit their messages simultaneously to $\mathrm{R}$ with transmit power $P_A$ and $P_B$, respectively.

In the broadcast phase, relay $\mathrm{R}$ employs linear processing with amplification matrix ${\bf{W}}$ to process the received signals and broadcast the processed signals to the nodes with harvested power $Q$. We assume that the source nodes $\rm S_{A}$ and $\rm S_{B}$ are equipped with a pair of transmitter-receiver antennas for signal transmission and reception, respectively. We denote the numbers of transmit and receive antennas at $\mathrm{R}$ as $M_T$ and $M_R,$ respectively. We use ${\bf h}_{XR} \in \mathbb{C}^{M_R \times 1}$ and ${\bf h}_{RX} \in \mathbb{C}^{M_T \times 1}$ to denote the directional channel vectors between the source node $X$'s $\in ({\rm S_A, S_B}$) transmit antennas to $\mathrm{R}$'s receive antennas, respectively, and that between the relay's transmit antenna(s) to the node $X$'s receiver antennas. The simultaneous transmission and reception of signals at the nodes in FD produces SI which if not properly handled inhibits the overall performance of FD systems. We thus consider using existing SI cancellation mechanisms (e.g., antenna isolation, digital and analog cancellation, etc.) to mitigate the SI. Due to imperfect channel estimation however, the SI cannot be completely eliminated \cite{single_channel}. We denote $h_{AA},$ $h_{BB},$ and ${\bf H}_{RR}\in \mathbb{C}^{M_R \times M_T}$ as the SI channel at the corresponding nodes. For simplicity, the residual SI (RSI) channel is modelled as Gaussian random variables with zero mean and variance $\sigma^2_X$, for $X\in \{{\rm S_A}, {\rm S_B}, {\rm R}\}$ \cite{single_channel}.

We further assume that $\rm R$ is equipped with a PS device which splits the received signal power at the relay such that a $\rho \in (0,1)$ portion of the received signal power is fed to the information receiver (IR) and the remaining $(1-\rho)$ portion of the power is fed to the energy receiver (ER) at the relay.

When the source nodes transmit their signals to the relay, it is known that the AF relay will incur a short delay to perform linear processing. It is generally assumed that the processing delay at the relay which corresponds to the processing time to implement FD operation is given by a $\tau$ symbol duration, which typically takes integer values \cite{delay}. The duration of $\tau$ is assumed to be short enough when compared to a time slot which has a larger number of data symbols, and therefore its affect on the achievable rate can be safely neglected \cite{delay}.

At time instant $n,$ the received signal ${\bf y}_{\rm R}[n]$ and the transmit signal ${\bf x}_{\rm R}[n]$ at the relay can be, respectively, written as
\begin{align}
{\bf y}_R[n] & = {\bf h}_{AR}s_A[n] + {\bf h}_{BR}s_B[n]+ {\bf H}_{RR}{\bf x}_R[n] + {\bf n}_R[n],\label{eqn:yR}\\
{\bf x}_{R}[n] &={\bf W}{\bf y}^{\rm IR}_R [n-\tau], \label{beam}
\end{align}
where ${\bf n}_R$ is the additive white Gaussian noise (AWGN) and ${\bf y}^{\rm IR}_R[n]$ is the signal split to the IR at $\mathrm{R}$ given by
\begin{equation}\label{receive_relay1}
{\bf y}^{\rm IR}_R[n] =\sqrt{\rho} {\bf y}_R[n],
\end{equation}
Accordingly, the relay output can be written as
\begin{equation}\label{beam2}
{\bf x}_R[n] =\sqrt{\rho} {\bf W}{\bf y}_R[n-\tau].
\end{equation}
Results in \cite{delay} showed that the capacity of relay networks with delay depends only on the relative path delays from the sender to the receiver and not on absolute delays. Accordingly, after using (\ref{eqn:yR})--(\ref{beam2}) recursively, the overall relay output can be written as \cite{delay}
\begin{multline}\label{beam3}
{\bf x}_R[n]={\bf W} \sum^{\infty}_{j = 0} ({\bf H}_{RR} {\bf W})^j \Big[ \sqrt{\rho} \big({\bf h}_{AR}s_A[n-j\tau-\tau]\\
+ {\bf h}_{BR}s_B[n-j\tau-\tau]+ {\bf n}_R[n-j\tau-\tau]\big)\Big],
\end{multline}
where $j$ denotes the index of the delayed symbols. Its covariance matrix can be expressed as
\begin{multline}\label{cov}
\mathbb{E}[{\bf x}_R {\bf x}^{\dagger}_R]=\\
\rho\Bigg[P_A{\bf W}\sum^{\infty}_{j=0} ({\bf H}_{RR} \mathrm{\bf W})^j {\bf h}_{AR}{\bf h}_{AR}^\dagger (( {\bf H}_{RR} {\bf W} )^j)^\dagger {\bf W}^\dagger \\
+P_B  {\bf W}\sum^{\infty}_{j=0} ({\bf H}_{RR} {\bf W})^j {\bf h}_{BR}{\bf h}_{BR}^\dagger (( {\bf H}_{RR} {\bf W} )^j)^\dagger {\bf W}^\dagger\\
+{\bf W} \sum^{\infty}_{j=0}( {\bf H}_{RR} {\bf W} {\bf W}^\dagger {\bf H}_{RR}^\dagger )^j {\bf W}^\dagger\Bigg].
\end{multline}
The relay's transmit covariance is a very complicated function of ${\bf W}.$ To keep the optimization problem more tractable, we add the zero-forcing (ZF) solution constraint such that the optimization of ${\bf W}$ nulls out the RSI from the relay output to the relay input \cite{Alex_2}. To this end, it is easy to check from (\ref{beam3}) that the condition below is sufficient:
\begin{equation}
{\bf WH}_{RR} {\bf W} = 0.
\end{equation}
Accordingly, (\ref{beam3}) becomes
\begin{multline}
{\bf x}_R[n]={\bf W} [ \sqrt{\rho} ({\bf h}_{AR}s_A[n-\tau]
 + {\bf h}_{BR}s_B[n-\tau]\\
+{\bf n}_R[n-\tau])]
\end{multline}
with the covariance matrix
\begin{multline}
\mathbb{E}[{\bf x}_R {\bf x}^{\dagger}_R] = \rho P_A {\bf W}{\bf h}_{AR}{\bf h}_{AR}^\dagger {\bf W}^\dagger\\
+ \rho P_B {\bf W}{\bf h}_{BR}{\bf h}_{BR}^\dagger{\bf W}^\dagger
+\rho{\bf W}{\bf W}^\dagger.
\end{multline}
The relay output power is given as
\begin{align}
P_R &= \mathrm{trace} ( \mathbb{E}[{\bf x}_R {\bf x}^{\dagger}_R])\notag\\
&= \rho\left[ P_A \|{\bf W}{h}_{AR} \|^2 + P_B \|{\bf W}{\bf h}_{BR}\|^2 + \mathrm{trace} ({\bf W}{\bf W}^\dagger)\right].
\end{align}
In the second time slot, the received signal at $\rm S_A$ is given as
\begin{align}
y_A[n] &={\bf h}^{\dagger}_{RA}{\bf x}_R[n] +  {h}_{AA}s_A[n] + n_A[n]\nonumber\\
&=\sqrt{\rho}\Big( {\bf h}^{\dagger}_{RA} {\bf W}{\bf h}_{AR} s_A[n-\tau] +{\bf h}^{\dagger}_{RA} {\bf W}{\bf h}_{BR} s_B[n-\tau] \nonumber\\
&\quad\quad\quad+{\bf h}^{\dagger}_{RA} {\bf W}{\bf n}_R[n]\Big)+  {h}_{AA}s_A[n] + n_A[n].\label{signode1}
\end{align}
After cancelling its own signal $s_A[n - \tau],$ (\ref{signode1}) becomes
\begin{multline}\label{signode2}
y_A[n]= \sqrt{\rho}\left({\bf h}^{\dagger}_{RA} {\bf W}{\bf h}_{BR} s_B[n-\tau]+{\bf h}^{\dagger}_{RA}{\bf W}{\bf n}_R[n]\right)\\
+ {h}_{AA}s_A[n]+  n_A[n].
\end{multline}
The received signal-to-interference-plus-noise ratio (SINR) at $\rm S_A$, denoted by $\gamma_A$, can be expressed as
\begin{equation}
\gamma_{A} = \frac{\rho P_B |{\bf h}^{\dagger}_{RA} {\bf W} \mathbf{h}_{BR}|^2}{ \rho\|\mathbf{h}^{\dagger}_{RA} \mathbf{W}\|^2+ P_A| {h_{AA}}|^2 + 1}.
\end{equation}
Similarly, the received SINR at $\rm S_B$ can be written as
\begin{equation}
\gamma_{B} = \frac{\rho P_A |\mathbf{h}^{\dagger}_{RB} \mathbf{W} \mathbf{h}_{AR}|^2}{ \rho\|\mathbf{h}^{\dagger}_{RB} \mathbf{W}\|^2+ P_B| {h_{BB}}|^2 + 1}.
\end{equation}
The corresponding rates are then given by $R_{A} = \log_2(1+\gamma_A)$ and $R_{B} = \log_2(1+\gamma_B)$ at $\rm S_A$ and $\rm S_B$, respectively. Meanwhile, the signal split to the ER at the relay can be written as
\begin{multline}\label{receive_relay}
{\bf y}^{\rm ER}_R[n] =\sqrt{1-\rho}\times\\
({\bf h}_{AR}s_A[n] + {\bf h}_{BR}s_B[n]+{\bf H}_{RR}{\bf x}_R[n]+{\bf n}_R[n]).
\end{multline}
The harvested energy is thus given by \cite{ruhuul}
\begin{equation}
Q = \beta({1-\rho})\left(|{\bf h}_{AR}|^2P_A + |{\bf h}_{BR}|^2P_B + \mathrm{\bar{E}} + M_T\right),
\end{equation}
in which $\mathrm{\bar{E}} = \mathbb{E}[{\bf x}_R{\bf x}^{\dagger}_R]$ and $\beta$ denotes the energy conversion efficiency of the ER at the relay which accounts for the loss in energy transducer for converting the RF energy to electrical energy. For simplicity, we assume $\beta = 1$ in this paper.

\subsection{Problem Statement}
Traditionally, in conventional HD relaying communications, bidirectional information exchange between $\rm S_A$ and $\rm S_B$ occur in two phases. In contrast, FD relaying systems reduce the entire operation to only one phase, hence increasing spectral efficiency. However, FD operation generates SI at each node, thus $\rm S_A$ and $\rm S_B$ may not always use their maximum transmit power as it increases the level of RSI, so each node must carefully choose its transmit power.

To ensure continuous information transfer between the two sources, the harvested energy at $\rm R$ must be above certain useful level, given by a prescribed threshold. As such, we formulate the following joint relay beamforming, transmit power and receive PS ratio ($0\le\rho\le 1$) optimization problem:
\begin{eqnarray}
  \max_{{\mathbf{W},\rho, P_A, P_B}}  \!\!\!& &\!\!\! R \equiv R_{A} + R_{B} \nonumber\\
  {\rm s.t.}   \!\!\!& &\!\!\!   Q\geq \bar{Q}, \quad P_R \leq P_{\max}, \quad \rho \in(0,1)\nonumber\\
\!\!\!& &\!\!\! P_A \leq P_{\max}, \quad P_B \leq P_{\max},\label{Problem}
\end{eqnarray}
where $R$ is the sum-rate of the FD AF relay SWIPT system, $P_R$ is the maximum transmit power at the relay and $\bar{Q}$ is the minimum amount of harvested energy required.

\section{Proposed Algorithm}
We consider single data stream only transmissions at the sources and assume that network coding principle encourages mixing of the data from the sources. Therefore, we can decompose $\mathbf{W}$ as $\mathbf{W} = {\mathbf{w}}_t \mathbf{w}^{\dagger}_r,$ where $\mathbf{w}_t$ is the transmit beamforming vector and $\mathbf{w}_r$ is the receive beamforming vector at the relay. Accordingly, the zeroforcing constraint can then be simplified to $(\mathbf{w}^{\dagger}_r {\bf H}_{RR} \mathbf{w}_t ) \mathbf{W} = 0$ or equivalently $(\mathbf{w}^{\dagger}_r {\bf H}_{RR} \mathbf{w}_t )=0$ since in general $\mathbf{W} \neq {\bf 0}$ \cite{Alex_2}. Also, without loss of optimality, we assume that $\|\mathbf{w}_r \|=1.$ The optimization problem in (\ref{Problem}) can therefore be re-expressed as
\begin{eqnarray}
 \max_{{\mathbf{w}_r,\mathbf{w}_t,  \rho, P_A, P_B}} \!\!\!& &\!\!\!
 \log_2 \left(1 +    \frac{\rho P_B C_{rB} |\mathbf{h}^{\dagger}_{RA} \mathbf{w}_t|^2}{ \rho\|\mathbf{h}^{\dagger}_{RA} \mathbf{w}_t \|^2 + P_A| {h_{AA}}|^2 + 1} \right)   \nonumber \\ 
   \!\!\!& + &\!\!\!    \log_2 \left(1 +  \frac{\rho P_A C_{rA} |\mathbf{h}^{\dagger}_{RB} \mathbf{w}_t|^2}{ \rho\|\mathbf{h}^{\dagger}_{RB} \mathbf{w}_t\|^2
+ P_B| {h_{BB}}|^2 + 1}\right) \nonumber\\ 
  {\rm s.t.} \nonumber\\
\!\!\!& &\!\!\!   ({1-\rho})(|\mathbf{h}_{AR}|^2P_A + |\mathbf{h}_{BR}|^2P_B \nonumber\\
 \!\!\!& + &\!\!\! \mathrm{\bar{E}} + M_T) \geq \bar{Q}   \nonumber \\
\!\!\!& &\!\!\!  \rho (P_A \|\mathbf{w}_t\|^2C_{rA} + P_B \|\mathbf{w}_t\|^2C_{rB} \nonumber\\
   \!\!\!& + &\!\!\!  \|\mathbf {w}_t\|^2 ) \leq P_R  \nonumber\\
\!\!\!& &\!\!\!  P_A \leq P_{\max}, P_B \leq P_{\max}  \nonumber\\
 \!\!\!& &\!\!\! \mathbf{w}^{\dagger}_r\mathbf{H}_{RR}\mathbf{w}_t=0,
 \label{y6}
 \end{eqnarray}
where $C_{rA}\triangleq |\mathbf{w}^{\dagger}_r \mathbf{h}_{AR}|^2$ and $C_{rB}\triangleq |\mathbf{w}^{\dagger}_r \mathbf{h}_{BR}|^2$.

\subsection{Parametrization of the Receive Beamforming Vector $\mathbf{w}_r$ }
Let us proceed to parametrize the receive beamforming vector $\mathbf{w}_r.$ We observe from (\ref{y6}) that  $\mathbf{w}_r$ is mainly involved in $|\mathbf{w}^{\dagger}_r\mathbf{h}_{AR}|^2$ and $|\mathbf{w}^{\dagger}_r\mathbf{h}_{BR}|^2$, so it has to balance the signals received from $\rm S_A$ and $\rm S_B$. Results obtained in \cite{complete} showed that $\mathbf{w}_r$ can be parameterized by $0 \leq \alpha \leq 1$ as
\begin{equation}
\mathbf{w}_r = \alpha \frac{\Pi_{\mathbf{h}_{BR}}\mathbf{h}_{AR}}{\|\Pi_{\mathbf{h}_{BR}}\mathbf{h}_{AR}\|} + \sqrt{1-\alpha}\frac{\Pi^{\perp}_{\mathbf{h}_{BR}}\mathbf{h}_{AR}}{\|\Pi^{\perp}_{\mathbf{h}_{BR}}\mathbf{h}_{AR}\|}.\label{pramet}
\end{equation}
It is worth noting that (\ref{pramet}) does not completely characterizes $\mathbf{w}_r$ as it is also involved in the ZF constraint $\mathbf{w}^{\dagger}_r\mathbf{H}_{RR}\mathbf{w}_t=0,$ however, this parameterization makes the problem more tractable. Thus, for given value of $\alpha,$ we can optimize $\mathbf{w}_t$ for known transmit powers $P_A$ and $P_B$ and for a fixed PS ratio ($\rho$). Consequently, we can perform a one-dimensional (1-D) search to find the optimal $\alpha^*$ which guarantees the optimal value for $\mathbf{w}_r.$

\subsection{Optimization of the Receiver PS ($\rho$) } \label{subsecR}
For fixed transmit power ($P_A$ and $P_B$) at $\rm S_A$ and $\rm S_B$, and given $\mathbf{w}_r$ and $\mathbf{w}_t$, the optimal receive PS ratio $\rho$ can be found. First, applying the concept of monotonicity between SINR and the rate, (\ref{y6}) can be rewritten as
 \begin{subequations}\label{rho}
\begin{eqnarray}
 \max_{{  \rho \in(0,1)}}  \!\!\!& &\!\!\!   \left(    \frac{\rho P_B C_{rB} |\mathbf{h}^{\dagger}_{RA} \mathbf{w}_t|^2}{ \rho\|\mathbf{h}^{\dagger}_{RA} \mathbf{w}_t \|^2 + P_A| {h_{AA}}|^2 + 1} \right) \nonumber \\ 
   \!\!\!&+ &\!\!\!   \left( \frac{\rho P_A C_{rA} |\mathbf{h}^{\dagger}_{RB} \mathbf{w}_t|^2}{ \rho\|\mathbf{h}^{\dagger}_{RB} \mathbf{w}_t\|^2
+ P_B| {h_{BB}}|^2 + 1}\right)\label{rho_obj}\\ 
   {\rm s.t.}  \!\!\!& &\!\!\!  ({1-\rho})(|\mathbf{h}_{AR}|^2P_A + |\mathbf{h}_{BR}|^2P_B  \nonumber\\
   \!\!\!&+ &\!\!\! \mathrm{\bar{E}} + M_T) \!\geq \! \bar{Q} \label{rhoc1}  \\
 \!\!\!& &\!\!\!  \rho (P_A \|\mathbf{w}_t\|^2C_{rA} + P_B \|\mathbf{w}_t\|^2C_{rB}  \nonumber\\
   \!\!\!&+ &\!\!\!  \|\mathbf {w}_t\|^2 ) \leq P_R. \label{rhoc2}
\end{eqnarray}
\end{subequations}
As shown in ($\ref{rho}$), we can easily verify that the objective function is an increasing function of $\rho.$ Hence, the optimal receive PS ratio ($\rho^*$) can be determined based on constraints (\ref{rhoc1}) and (\ref{rhoc2}) only. The optimal point will be the largest $\rho$ that satisfies both constraints. It is worth noting that the left hand side of constraint (\ref{rhoc1}) is a decreasing function of $\rho$ whereas that of constraint (\ref{rhoc2}) is an increasing function of $\rho.$ Now, the largest $\rho$ satisfying (\ref{rhoc1}) to equality is given by
  \begin{equation}
 \rho_l = 1 - \frac{\bar{Q}}{|\mathbf{h}_{AR}|^2{P_A} + |\mathbf{h}_{BR}|^2{P_B} + \mathrm{\bar{E}} + M_T}.
\end{equation}
On the other hand, the maximal $\rho$ satisfying constraint \eqref{rhoc2} to
equality is given by
\begin{equation}
\rho_m = \frac{P_R }{P_A \|\mathbf{w}_t\|^2C_{rA} + P_B \|\mathbf{w}_t\|^2C_{rB} + \|\mathbf{w}_t\|^2} \\.
\end{equation}
We investigate whether $\rho_l$ satisfies the constraint \eqref{rhoc2}. If it does, then it is the optimal solution $\rho^*.$ Otherwise, we perform a 1-D search over $\rho$ starting from $\rho_l$ until \eqref{rhoc2} is satisfied.  Clearly, if $\rho_m > \rho_l,$ then problem (\ref{rho}) becomes infeasible.

\subsection{Optimization of the Beamforming Vectors ($\mathbf{w}_t$ and $\mathbf{w}_r$)}
For fixed values of the receive PS ($\rho$) at the relay and the transmit power ($P_A$ and $P_B$) at the sources, problem (\ref{y6}) can be reformulated as
\begin{eqnarray}
 \max_{{\mathbf{w}_r,\mathbf{w}_t}} \!\!\!& &\!\!\!
 \log_2 \left(1 +    \frac{\rho P_B C_{rB} |\mathbf{h}^{\dagger}_{RA} \mathbf{w}_t|^2}{ \rho\|\mathbf{h}^{\dagger}_{RA} \mathbf{w}_t \|^2 + P_A| {h_{AA}}|^2 + 1} \right)   \nonumber \\ 
   \!\!\!& + &\!\!\!    \log_2 \left(1 +  \frac{\rho P_A C_{rA} |\mathbf{h}^{\dagger}_{RB} \mathbf{w}_t|^2}{ \rho\|\mathbf{h}^{\dagger}_{RB} \mathbf{w}_t\|^2
+ P_B| {h_{BB}}|^2 + 1}\right) \nonumber\\ 
  {\rm s.t.} \nonumber\\
\!\!\!& &\!\!\!   ({1-\rho})(|\mathbf{h}_{AR}|^2P_A + |\mathbf{h}_{BR}|^2P_B \nonumber\\
 \!\!\!& + &\!\!\! \mathrm{\bar{E}} + M_T) \geq \bar{Q}   \nonumber \\
\!\!\!& &\!\!\!  \rho (P_A \|\mathbf{w}_t\|^2C_{rA} + P_B \|\mathbf{w}_t\|^2C_{rB} \nonumber\\
   \!\!\!& + &\!\!\!  \|\mathbf {w}_t\|^2 ) \leq P_R  \nonumber\\
\!\!\!& &\!\!\!  P_A \leq P_{\max}, P_B \leq P_{\max}  \nonumber\\
 \!\!\!& &\!\!\! \mathbf{w}^{\dagger}_r\mathbf{H}_{RR}\mathbf{w}_t=0.
 \label{y6b}
 \end{eqnarray}
Let us proceed to separately optimize $\mathbf{w}_t$ and $\mathbf{w}_r$ in the next two subsections.

\subsection{Optimization of the Transmit Beamforming Vector ($\mathbf{w}_t$)}
Here, we first investigate how $\mathbf{w}_t$ is optimized  for given $\rho$ assuming the sources' transmit power ($P_A,$ $P_B$) are fixed. For convenience, we denote a semidefinite matrix  $\mathbf{W}_t \triangleq \mathbf{w}_t\mathbf{w}^{\dagger}_t.$ The optimization problem in (\ref{y6}) can be reformulated as
\begin{eqnarray}
 \max_{{\mathbf{W}_t\succeq0}} \!\!\!& &\!\!\! F(\mathbf{W}_t) \nonumber\\
  {\rm s.t.}
  \!\!\!& &\!\!\! \mathrm{trace} (\mathbf{W}_t) \leq \frac{P_R}{\rho(P_A C_{rA}+ P_B C_{rB}+ 1)}\nonumber\\
  \!\!\!& &\!\!\! (1-\rho)(|\mathbf{h}_{AR}|^2P_A + |\mathbf{h}_{BR}|^2P_B +  \mathrm{\bar{E}} + 1)\nonumber\\
\!\!\!& &\!\!\!   \geq \bar{Q}\nonumber\\
 \!\!\!& &\!\!\!   \mathrm{trace} (\mathbf{W}_t\mathbf{H}^{\dagger}_{RR} \mathbf{w}_r\mathbf{w}^{\dagger}_r\mathbf{H}_{RR}) = 0 \nonumber\\
\!\!\!& &\!\!\!  \mathrm{rank}(\mathbf{W}_t) = 1,\label{max1}
\end{eqnarray}
where $F(\mathbf{W}_t)$ is given as
\begin{multline}\label{FW}
F(\mathbf{W}_t) \triangleq\\
 \log_2\left( 1+ \frac{\rho P_B C_{rB}\mathrm{trace}(\mathbf{W}_t\mathbf{h}_{RA} \mathbf{h}^{\dagger}_{RA} ) }{\rho \mathrm{trace}(\mathbf{W}_t \mathbf{h}_{RA} \mathbf{h}^{\dagger}_{RA} ) + P_A| {h_{AA}}|^2 + 1}\right)\\
+\log_2 \left( 1 +  \frac{\rho P_A C_{rA} \mathrm{trace}( \mathbf{W}_t\mathbf{h}_{RB}   \mathbf{h}^{\dagger}_{RB} ) }{\rho \mathrm{trace}(\mathbf{W}_t \mathbf{h}_{RB} \mathbf{h}^{\dagger}_{RB} ) + P_B| {h_{BB}}|^2 + 1}\right).
\end{multline}
Obviously, (\ref{FW}) is not a concave function, making it difficult to solve. To solve (\ref{FW}), we propose to use the difference of convex programming (DC) to find the local optimum point. Therefore, we can express $F(\mathbf{W}_t)$ as a difference of two concave functions denoted $f(\mathbf{W}_t)$ and $g(\mathbf{W}_t)$, i.e.,
\begin{eqnarray}
 F(\mathbf{W}_t)\!\!\!&=&\!\!\!
 \log_2((\rho P_B C_{rB}+ \rho)\mathrm{trace}(\mathbf{W}_t\mathbf{h}_{RA}\mathbf{h}^{\dagger}_{RA} )\nonumber\\
\!\!\!&+&\!\!\! P_A |{h_{AA}}|^2 + 1) - \log_2(\rho \mathrm{trace}(\mathbf{W}_t\mathbf{h}_{RA}\mathbf{h}^{\dagger}_{RA})\nonumber\\
\!\!\!&+&\!\!\!   P_A |{h_{AA}}|^2 + 1)\nonumber\\
\!\!\!& &\!\!\! + \log_2((\rho P_A C_{rA}+ \rho)\mathrm{trace}(\mathbf{W}_t\mathbf{h}_{RB}\mathbf{h}^{\dagger}_{RB} )\nonumber\\
\!\!\!&+&\!\!\! P_B |{h_{BB}}|^2 + 1) - \log_2(\rho \mathrm{trace}(\mathbf{W}_t\mathbf{h}_{RB}\mathbf{h}^{\dagger}_{RB}) \nonumber\\
\!\!\!&+&\!\!\!  P_B |{h_{BB}}|^2 + 1)\nonumber\\
\!\!\!& = &\!\!\! f(\mathbf{W}_t)- g(\mathbf{W}_t),
\end{eqnarray}
where
\begin{eqnarray}
f(\mathbf{W}_t) \!\!\!& \triangleq &\!\!\! \log_2((\rho P_B C_{rB}+ \rho )\mathrm{trace}(\mathbf{W}_t\mathbf{h}_{RA}\mathbf{h}^{\dagger}_{RA} )\nonumber\\
\!\!\!&+&\!\!\! P_A |{h_{AA}}|^2 + 1)+ \log_2((\rho P_A C_{rA}+ \rho )\nonumber\\
\!\!\!&\times&\!\!\!\mathrm{trace}(\mathbf{W}_t\mathbf{h}_{RB}\mathbf{h}^{\dagger}_{RB} ) + P_B |{h_{BB}}|^2 + 1),
\label{fw_t}
\end{eqnarray}
\begin{eqnarray}
g(\mathbf{W}_t)\!\!\!\!& \triangleq &\!\!\!\!\!  \log_2(\rho \mathrm{trace}(\mathbf{W}_t\mathbf{h}_{RA}\mathbf{h}^{\dagger}_{RA}) + P_A |{h_{AA}}|^2 + 1) \nonumber\\
\!\!\!&+&\!\!\!  \log_2(\rho \mathrm{trace}(\mathbf{W}_t\mathbf{h}_{RB}\mathbf{h}^{\dagger}_{RB}) + P_B |{h_{BB}}|^2 + 1). \nonumber\\
\label{gw_t}
\end{eqnarray}
Clearly, we notice that $f(\mathbf{W}_t)$ is a concave function while $g(\mathbf{W}_t)$ is a convex function. The adopted solution approach is to approximate $g(\mathbf{W}_t)$ by a linear function. As a result, we define the first order approximation of $g(\mathbf{W}_t)$ around the point $f(\mathbf{W}_{t,k})$ as given in (\ref{lineartn}) (see top of next page). Thus, we exploit the concept of DC programming to sequentially solve the resulting convex problem:
\begin{figure*}[!t]
\normalsize
\begin{eqnarray}
g_L(\mathbf{W}_t; \mathbf{W}_{t,k}) \!\!\!&=&\!\!\!
 \frac{1}{\mathrm{ln}(2)} \frac{\rho\mathrm{trace}{((\mathbf{W}_t-\mathbf{W}_{t,k})\mathbf{h}_{RA}\mathbf{h}^{\dagger}_{RA})}+ \mathrm{trace}((\mathbf{W}_t-\mathbf{W}_{t,k})\mathbf{h}_{RA}\mathbf{h}^{\dagger}_{RA})}{\rho \mathrm{trace}(\mathbf{W}_{t,k}\mathbf{h}_{RA}\mathbf{h}^{\dagger}_{RA})+ P_A|{h_{AA}}|^2+1}\nonumber\\
 \!\!\!&+&\!\!\! \frac{1}{\mathrm{ln}(2)} \frac{\rho\mathrm{trace}{((\mathbf{W}_t-\mathbf{W}_{t,k})\mathbf{h}_{RB}\mathbf{h}^{\dagger}_{RB})}+ \mathrm{trace}((\mathbf{W}_t-\mathbf{W}_{t,k})\mathbf{h}_{RB}\mathbf{h}^{\dagger}_{RB})}{\rho \mathrm{trace}(\mathbf{W}_{t,k}\mathbf{h}_{RB}\mathbf{h}^{\dagger}_{RB})+ P_B|{h_{BB}}|^2+1}\nonumber\\
 \!\!\!&+&\!\!\! \log_2(\rho \mathrm{trace}(\mathbf{W}_{t,k}\mathbf{h}_{RA}\mathbf{h}^{\dagger}_{RA}) + P_A |{h_{AA}}|^2+1) +\log_2(\rho \mathrm{trace}(\mathbf{W}_{t,k}\mathbf{h}_{RB}\mathbf{h}^{\dagger}_{RB}) +  P_B |{h_{BB}}|^2 + 1).\label{lineartn}
\end{eqnarray}
\hrulefill
\end{figure*}
\begin{eqnarray}
\mathbf{W}_{t,k+1} \!\!\!&=&\!\!\! \mbox{arg} \max_{{\mathbf{W_t}}} f(\mathbf{W}_t) - g_L (\mathbf{W}_t; \mathbf{W}_{t,k}) \nonumber\\ {\rm s.t.}
\!\!\!& &\!\!\! \mathrm{trace}(\mathbf{W}_t) = \frac{P_R}{\rho(P_A C_{rA}+ P_B C_{rB}+ 1)} \nonumber\\
 \!\!\!& &\!\!\! (1-\rho)(|\mathbf{h}_{AR}|^2P_A + |\mathbf{h}_{BR}|^2P_B \nonumber\\
 \!\!\!&+ &\!\!\!  \mathrm{\bar{E}} + 1) \geq \bar{Q}\nonumber\\
  \!\!\!& &\!\!\! \mathrm{trace}(\mathbf{W}_t\mathbf{H}^{\dagger}_{RR}\mathbf{w}_r\mathbf{w}^{\dagger}_r\mathbf{H}_{RR})=0.\label{main_prob2}
\end{eqnarray}
Now, (\ref{max1}) can be solved by (i) choosing an initial point $\mathbf{W}_t$ and (ii) for $k= 0,1,\dots$, solving (\ref{main_prob2}) until convergence. The rank-1 constraint associated with solving (\ref{main_prob2}) is guaranteed by the results in \cite[Theorem 2]{new_result} when $M_T > 2$. As a consequence, we ignore the rank-1 constraint on  $\mathbf{W}_t.$ The decomposition of $\mathbf{W}_t$ leads to the optimal solution $\mathbf{w}^{\dagger}_t.$

\subsection{Optimization of the Receive Beamforming Vector ($\mathbf{w}_r$)}
Given $\mathbf{w}_t,$ the value of the optimal $\mathbf{w}_r$ can be obtained by performing a 1-Dimensional search on $\alpha$ to find the maximum $\alpha^*$ which maximizes the sum-rate $R(\mathbf{w}_r)$ for given values of $\rho \in (0,1),$ $P_A$ and $P_B.$ Algorithm \ref{algorithm1} summarizes this procedure. The lower bound of the rate search denoted $(R_{A}+ R_{B})_{\rm low}$ is evidently zero while the upper bound $(R_{A}+ R_{B})_{\max}$ in contrast, denotes the achievable sum-rate at zero RSI. As a consequence, if we know optimal $\alpha^*,$ then the optimal $\mathbf{w}^*_r$ can be obtained from (\ref{pramet}).
 \begin{algorithm}
 \caption{Procedure for solving problem  (\ref{y6b})} 
\label{algorithm1}
\begin{algorithmic}[1]
\State Set $0 \le \alpha \le 1,$ $0 \le \rho \le 1,$ $P_A > 0$ and $P_B > 0$ as non-negative real-valued scaler and obtain $\mathbf{w}_r$ as given in (\ref{pramet}).

\State At step $k,$ set $\alpha(k) = \alpha(k - 1) + \triangle\alpha$ until $\alpha(k)=1,$ where $\triangle \alpha$ is the searching step size.

 \State Initialize $(R_{A} + R_{B})_{\rm low} = 0$ and $(R_{A} + R_{B})_{\rm up} = (R_{A}
  + R_{B})_{\max}$.

\State \textbf{Repeat}

 		 a) Set $R \leftarrow \frac{1}{2}((R_{A} + R_{B})_{\rm low} +  (R_{A} + R_{B})_{\rm up})$
 		
 		b) Obtain the optimal relay transmit beamforming

        vector $\mathbf{w}_t$ by solving problem (\ref{main_prob2}).
 		
		c) Using the bisection search method, update
		
		the value of $R$: if (b) is feasible, set
		
		$(R_{A} + R_{B})_{\rm low} = R$; otherwise, $(R_{A} + R_{B})_{\rm up} = R$.
		
 \State \textbf{Until}
 $(R_{A} + R_{B})_{\rm up} - (R_{A} + R_{B})_{\rm low} < \epsilon,$ where $\epsilon$ is a small positive number. Consequently, we get $R(\alpha(k)).$
	
\State $k = k+1$

\State Find the optimal $\alpha^*$ by comparing all $R(\alpha(k))$ that yields maximal $R$. Corresponding $\mathbf{w}_t$ is the optimal one.
\State Obtain the optimal $\mathbf{w}_r^*$ from (\ref{pramet}) using $\alpha^*$.
\end{algorithmic}
\end{algorithm}

\subsection{Optimization of the Source Power ($P_A,$ $P_B$)} \label{subsecP}
A major concern towards achieving FD communication is the presence of SI resulting from a node's own transmit signal. In this work, we assume that each source node houses a transmitter-receiver pair for signal transmission and reception, respectively. As a result, they cannot suppress the RSI in the spatial domain and therefore they may not consistently use their full transmit power. In contrast, the relay is equipped with at least two transmit and receiver antennas, thus, it can completely cancel the resulting SI and can transmit with full power ($P_R$) \cite{joint}. In this subsection, we investigate the optimal power solution ($P_A,$ $P_B$) at sources $\rm S_A$ and $\rm S_B$, respectively, assuming $\mathbf{w}_t, $ $\mathbf{w}_r$ and $\rho$ all being fixed.

For convenience, we define $C_{At} \triangleq |\mathbf{h}^{\dagger}_{RA}\mathbf{w}_t|^2$ and $C_{Bt} \triangleq |\mathbf{h}^{\dagger}_{RB}\mathbf{w}_t|^2.$ It can be easily verified that at optimum, at least one source should achieve its maximum power \cite{joint,Transmit_strategies}, i.e., $P_A = P_{\max}$ or $P_B = P_{\max}.$ Exploiting this fact, we can relax (\ref{y6}) into two sub-problems: (i) $P_A = P_{\max}$, (ii)$P_B = P_{\max}.$ Then we solve each sub-problem individually. Since  case (i) and case (ii) are symmetric, we take case (i) as an example. The objective function of (\ref{y6}) is reformulated as
\begin{subequations}\label{Pb^*}
\begin{eqnarray}
 \max_{{P_B}}
  \!\!\!& &\!\!\! \log_2 \left(1+ \frac{\rho P_B C_{rB} C_{At}}{ \rho C_{At} + P_{\max}| {h_{AA}}|^2 + 1} \right)  \\
   \!\!\!& + &\!\!\! \log_2 \left(1+ \frac{\rho P_{\max} C_{rA} C_{Bt}}{ \rho C_{Bt} + P_B| {h_{BB}}|^2 + 1} \right) \label{PB_constrt^}\\
   {\rm s.t.}
  \!\!\!& &\!\!\!({1-\rho})(|\mathbf{h}_{AR}|^2P_{\max} + |\mathbf{h}_{BR}|^2P_B  \nonumber\\
  \!\!\!&+&\!\!\! \mathrm{\bar{E}} + M_T) \geq \bar{Q} \label{a} \\
\!\!\!& &\!\!\!  \rho (P_{\max} \|\mathbf{w}_t\|^2C_{rA} + P_B \|\mathbf{w}_t\|^2C_{rB}  \nonumber\\
 \!\!\!&+&\!\!\! \|\mathbf {w}_t\|^2 ) \leq P_R  \label{b}\\
\!\!\!& &\!\!\!  \mathbf{w}^{\dagger}_r\mathbf{H}_{RR}\mathbf{w}_t=0 \\
\!\!\!& &\!\!\!  P_B \leq P_{\max}.
 \label{power}
\end{eqnarray}
\end{subequations}
It is obvious that the objective function (\ref{Pb^*}) is an increasing function of $P_B.$ Hence, optimal $P^*_B$ can be determined based on the constraints (\ref{a}) and (\ref{b}) only. The optimal point will be the largest $P_B$ that satisfies both constraints. Now the smallest $P_B$ satisfying (\ref{a}) to equality is given by
\begin{equation}
P_{Bs} = \frac{\frac{\bar{Q}}{1-\rho} - |\mathbf{h}_{AR}|^2P_{\max} - \bar{E} - M_T}{|\mathbf{h}_{BR}|^2}.
\end{equation}
On the other hand, the maximal $P_B$ satisfying constraint (\ref{b}) to equality is given by
\begin{equation}
P_{Bm} = \frac{\frac{P_R}{\rho} - P_{\max} \| \mathbf{w}_t\|^2 C_{rA} - \| \mathbf{w}_t\|^2}{\| \mathbf{w}_t\|^2 C_{rB}}.
\end{equation}
We investigate whether $P_{Bm}$ satisfies the constraint (\ref{a}). If it does, then it is the optimal solution $P^*_B.$ Otherwise, we perform a 1-D search over $P_B$ starting from $P_{Bm}$  until $P_{Bs}$ is reached. Clearly, if $P_{Bs} > P_{Bm},$ then  (\ref{Pb^*}) becomes infeasible.

\subsection{Iterative Update}
The original sum-rate maximization problem in (\ref{y6}) can now be solved by an iterative technique shown in Algorithm \ref{alg2}. The objective function in (\ref{y6}) is continuously updated by Algorithm \ref{alg2} until convergence.
  \begin{algorithm}
\caption{Procedure for solving problem (\ref{y6})}
\label{alg2}
\begin{algorithmic}[1]
\State Initialize  $0 \le \rho \le 1$.
\State \textbf{Repeat}

		a) Obtain $\mathbf{w}_t^*$ and $\mathbf{w}_r^*$  for fixed  $P_A$  and $P_B$ using
		
		 Algorithm~\ref{algorithm1}.
		
		b) Obtain the optimal $\rho^*$ following the procedure in

        subsection~\ref{subsecR}
		
		c) Obtain the optimal $P^*_A$  and $P^*_B$  following the
		
		procedure in subsection~\ref{subsecP}
\State \textbf{Until} convergence.
\end{algorithmic}
\end{algorithm}

\section{Numerical Results}\label{num_1}
Here, we evaluate the performance of the proposed algorithm through computer simulations. Specifically, we consider a flat fading environment for the communication channel where the  fading coefficients are characterized as complex Gaussian numbers with zero mean which are independent and identically distributed (i.i.d.). The simulation results are averaged over $500$ independent channel realizations and the relay transmit power is given as $P_R = -5$ (dB).
\begin{figure}[ht]
\centering
\includegraphics*[width=5.8cm]{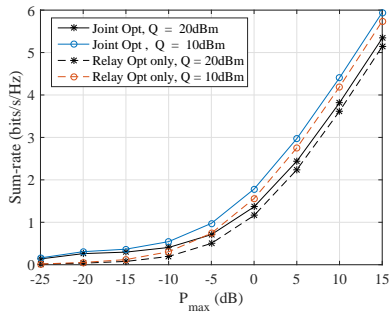}
\caption{Sum-rate versus $P_{\max}$.}
\label{Sec_R1}
\end{figure}

\begin{figure}[ht]
\centering
\includegraphics*[width=5.8cm]{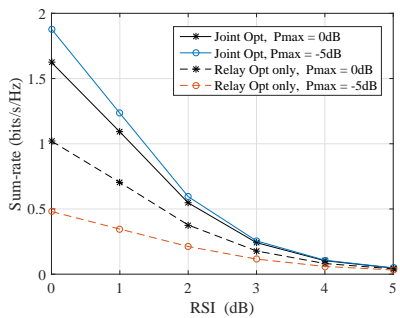}
\caption{Sum-rate versus RSI.}
\label{Sec_R2}
\end{figure}

In Fig.~\ref{Sec_R1}, we show the sum-rate results against the transmit power budget $P_{\max}$ (dB) for various harvested power constraint. The proposed scheme (`Joint Opt' in the figure) is compared with those of the relay optimization only (`Relay Opt only' in the figure), at optimal PS coefficient ($\rho^*$). Remarkably, the proposed scheme achieves higher sum-rate compared to the sum-rate of the Relay Opt only scheme which essentially shows the need for joint optimization. Also, as the harvested energy constraint decreases from $20$ dBm to $10$ dBm, there is an increase in the achievable sum-rate for both schemes. However, the joint optimization scheme achieves a higher sum-rate compared to the Relay Opt only scheme. 

Finally, in Fig. \ref{Sec_R2} the impact of the RSI on the sum-rate is investigated. To be specific, we analyse the performance of SWIPT in FD relay systems in terms of the sum-rate for both Relay Opt only and joint optimization versus the RSI (dB) for different values of transmit power constraint. As can be observed in Fig. \ref{Sec_R2}, as the RSI increases, there is a corresponding decrease in the achievable sum-rate.

\section{Conclusion}\label{conc_1}
In this paper, we investigated the joint transmit power and relay beamforming optimization for SWIPT in FD MIMO two-way AF relaying system and proposed an algorithm to maximize the achievable sum-rate subject to the total transmit power and harvested power constraints. Using DC and 1-D search, we achieved the joint optimization. Simulation results corroborates the importance of joint optimization.

\end{document}